# Spectral Evidence for an Inner Carbon-Rich Circumstellar Belt in the Young HD36546 A-Star System

C.M. Lisse[1], M.L. Sitko[2], R.W. Russell[3], M. Marengo[4], T. Currie[5], C. Melis[6]

T. Mittal[7], I. Song[8]

[1] JHU-APL, 11100 Johns Hopkins Road, Laurel, MD 20723 carey.lisse@jhuapl.edu, ron.vervack@jhuapl.edu

[2] Department of Physics, University of Cincinnati, Cincinnati, OH 45221-0011 and Space Science Institute, Boulder, CO 80301, USA  sitkoml@ucmail.uc.edu

[3] The Aerospace Corporation, Los Angeles, CA 90009, USA  ray.russell@aero.org

[4] Department of Physics and Astronomy, 12 Physics Hall, Iowa State University, Ames, IA 50010 mmarengo@iastate.edu

[5] Subaru Telescope, National Astronomical Observatory of Japan, National Institutes of Natural Sciences, Hilo, HI, USA  currie@naoj.org

[6] Center for Astrophysics and Space Sciences, University of California, San Diego, CA 92093-0424, USA  cmelis@ucsd.edu

[7] Department of Earth and Planetary Sciences, McCone Hall, University of California at Berkeley, Berkeley, CA 94720  tmittal2@berkeley.edu

[8] Department of Physics & Astronomy, University of Georgia, Athens, GA 30602-2451 USA  song@physast.uga.edu

# Abstract

Using the NASA/IRTF SpeX & BASS spectrometers we have obtained 0.7 - 13 µm observations of the newly imaged HD36546 debris disk system. The SpeX spectrum is most consistent with the photospheric emission expected from an $L_* \sim 20~L_\odot$, solar abundance A1.5V star with little to no extinction, and excess emission from circumstellar dust detectable beyond 4.5 µm. Non-detections of CO emission lines and accretion signatures point to the gas-poor circumstellar environment of a very old transition disk. Combining the SpeX + BASS spectra with archival WISE/AKARI/IRAS/Herschel photometry, we find an outer cold dust belt at ~135K and 20—40 AU from the primary, likely coincident with the disk imaged by Subaru (Currie *et al.* 2017), and a new second inner belt with temperature ~570K and an unusual, broad SED maximum in the 6-9 µm region, tracing dust at 1.1—2.2 AU. An SED maximum at 6-9 µm has been reported in just two other A-star systems, HD131488 and HD121191, both of ~10 Myr age (Melis *et al.* 2013). From *Spitzer*, we have also identified the ~12 Myr old A7V HD148567 system as having similar 5 - 35 µm excess spectral features (Mittal *et al.* 2015). The Spitzer data allows us to rule out water emission and rule in carbonaceous materials - organics, carbonates, SiC - as the source of the 6-9 µm excess. Assuming a common origin for the 4 young A-star systems' disks, we suggest they are experiencing an early era of carbon-rich planetesimal processing.

# 1. Introduction

HD 36546 hosts a bright, wide circumstellar disk that has just recently been imaged for the first time by Currie *et al.* (2017) using Subaru/SCExAO. Identified in the WISE all-sky survey as one of the most promising new debris disk detections (Padgett 2011; McDonald *et al.* 2012; Wu *et al.* 2013; Liu *et al.* 2014; Cotten & Song 2016), its extent (~2" in diameter) and brightness ($L_{excess}/L_*$ value of ~4 x $10^{-3}$) make it an important target for further study. Located slightly foreground (d = 114 pc, van Leeuwen 2007) to the 1–2 Myr old Taurus-Auriga star-forming region (d = 130 -160 pc, Kenyon *et al.* 2008; Luhman *et al.* 2009; Torres *et al.* 2009), it was not reported by IRAS in its all-sky survey (Aumann 1985) due to confusion with background Taurus Dark Cloud emission. Thus few characterizing measurements have been made of the system, even some 30+ years after the IRAS mission - mainly just Michigan survey, Tycho survey, WISE, and AKARI photometry (*Vizier* photometric database http://vizier.u-strasbg.fr/vizier/sed/; Pickles & Depagne 2010).

In this letter we present new 0.7 – 5.0 µm SpeX (Rayner *et al.* 2003, 2009; Vacca *et al.* 2003, 2004; Cushing *et al.* 2004) and 3-13 µm BASS (Hackwell *et al.* 1990) spectra of the HD36546 system taken as part of the 100+ hours Near InfraRed Disk Survey (NIRDS; Lisse *et al.* 2013, 2018; 50+ systems to date) at the NASA/IRTF 3m. Our spectra are novel in that they cover the entire wavelength range from 0.7 - 5.0 *µ*m at high spectral resolution and the 5-13 µm region *at all*, allowing us to characterize the primary star's photospheric absorption features and search for warm inner excess emission due to circumstellar gas and dust in the system. We also include new high SNR Herschel photometry in our analysis that extend system FIR measurements out to 160 µm.

# 2. Observations

We first observed HD36546 on 13 Jan 2017 UT from the NASA/IRTF 3m using SpeX. The instrument provided R = 2000 to 2500 observations from 0.7 – 5.0 µm in two orders, termed SXD (for "short cross-dispersed") and LXD (for "long cross-dispersed"), configured with an 0.3" slit (Rayner *et al.* 2003; 2009). Our observational setup was identical to that used for other NIRDS debris disk studies (e.g. Lisse *et al.* 2012, 2013, 2015, 2017). The nearby A0V star used in our Spextool data reduction as a calibration standard (Vacca *et al.* 2004) was HD34203 ($K$ = 5.46), picked to match the HD36546 primary ($K$ = 6.8) in color, effective temperature and spatial proximity (the best-fit

BtVtJHKs HD36546 photometry gave Teff = 9450K and (B-V)$_o$ = 0.07, consistent with an unreddened A2V star; Pecaut & Mamajek 2013). We observed HD36546 in SXD mode with a total on-target integration time of 960 sec and in LXD mode for 1800 sec, while the calibrator star HD 34203 was observed for 480 sec in SXD and 1440 sec in LXD mode. The instrument behavior was nominal, and the weather was excellent, with < 20% relative humidity at the summit and seeing ~0.6" at K-band, allowing for good sky correction and stellar calibration. Both stars were observed in ABBA nod mode to remove telescope and sky backgrounds. We observed HD36546 1.5 hrs after evening twilight and before meridian transit, with the position angle (P.A.) of the slit on the sky between 241 and 265$^o$ parallel to the long axis of the circumstellar disk, as described by Currie *et al.* (2017).

Using BASS (Hackwell *et al.* 1990), we obtained 3-13 μm, R = 30 - 125 follow-up observations of HD36546 at the IRTF on 29 Jan 2017 UT. Again the weather was very good, and the effective on-target time was ~15 min of integration. A N-S chopper throw of ~10.2" was utilized along with the ~4.2" wide BASS beam, which encompassed the system's whole disk. The nearby star used as a sky calibrator was HD29139 (α Tau). The median BASS SNR achieved in the 3.1-5.1 μm region of HD36546's spectrum was 54, and in the 8-13 μm region was 5.1.

## 3.     Results

Figure 1 shows our HD36546 SpeX measurements in the context of 6 other NIRDS main sequence early A-type stellar spectra. All of the spectra are as taken and calibrated vs. a nearby standard A0V star; no de-reddening has been applied. The typical spectral behavior for our program systems matches the photospheric model well from 0.7 - 1.3 μm, exceeds it slightly (if at all) from 1.3 - 3 μm, then matches it again from 3 - 5 μm. For a small subset of ~8 young NIRDS stars (e.g., HD113766, HD15407A, HD23514; Lisse *et al.* 2015; Dore *et al.* 2016), with abundant warm (200 - 500K) circumstellar dust, we find a close match to a stellar photospheric spectrum out to ~3 μm, then a rapid, exponential increase in excess from 3 to 5 μm that continues to rise into the large 8-13 μm warm silicate emission complex excesses seen in associated mid-infrared system spectra. The ~10 Myr old star HR4796A (A0.5V according to our NIRDS measurements, Lisse *et al.* 2017) shows an unusual excess that begins at ~1.5 μm and increases in a roughly linear fashion outwards to ~10 μm, where it is overtaken by rising thermal emission from cold (~100K) circumstellar material. The ~10 Myr old star HD131488 (A3V according to our NIRDS measurements; Melis *et al.* 2013 list it as an A1V) shows an unusual excess that begins at ~3.2 μm and is relatively flat outwards to 5 μm, then peaks at 6-7 μm, falling afterwards at 8-9 μm (Melis *et al.* 2013; Fig 2 inset). Of all our NIRDS stars, the coarse spectral behavior of HD 36546 in the near-infrared looks most similar to that of HD 131488, although with smaller excess that is just beginning to be seen at 4.5 - 5.0 μm in our SpeX data.

We performed a weighted $\chi^2$ spectral PHOENIX/NextGen model fit to HD36546's photosphere from 0.95 to 2.4 μm in (E(B-V), $T_{eff}$ and log g), with the model normalized to the 0.72 to 0.82 μm region of our SpeX spectrum. The best fit parameters E(B-V) = 0.01, $T_{eff}$ = 8920 ± 80 K, and log g = 4.50 ± 0.25, fall between the values for a solar abundance A1V and a solar abundance A2V star in E. Mamajek's stellar classification scheme (Pecaut & Mamajek 2013). For the given distance of 114 pc from Earth, the V=6.95 Tycho magnitude for HD36546 is consistent with $L_*$ ~ 20 $L_\odot$, which compares well with the Hipparcos L = 25 $L_\odot$ for A1V Sirius at 2.68 pc. The excellent model fit is consistent with WISE and NIR 2MASS and Tycho synthetic photometry (*Vizier* photometric database http://vizier.u-strasbg.fr/vizier/sed/; Pickles & Depagne 2010) of HD36546 except at the longest wavelengths, where circumstellar excess flux becomes important. Our 0.7 - 5.0 μm SpeX results for HD36546 are shown in detail in Figure 2 in black, in comparison to the best fit PHOENIX/NextGen photospheric model. There are no detectable HBrγ and CO emission lines due to fluorescing circumstellar gas or accretion, or HeI 1.083 μm, FeII 1.256/1.644 μm, and $H_2$ S(1) 1-0 2.121 μm emission lines due to strong wind outflow within the noise of the measurement (Connelley & Greene 2010, 2014). No emission line due to neutral PAHs is present at 3.29 μm. (These lines were easily detected in the NIRDS survey for the gas-rich A-star YSO 51 Oph and the bright Nova Oph, so we know that we are capable of detecting them with SpeX if they are present.)

Looking at the range of allowed models, we find that A0.5V - A2.5V models with E(B-V) ~ 0.01 are allowed by our data; models earlier than A0.5V require E(B-V) > 0.06 and very high values of log g (for a young A-star), and even then they produce a worse fit to the continuum between the hydrogen absorption lines. Low extinction photospheric models are consistent with HD36546 at d = 114 pc (van Leeuwen 2007) being foreground to the 1–2 Myr old Taurus-Auriga star-forming region at ~140 pc (Kenyon *et al.* 2008; Luhman *et al.* 2009; Torres *et al.* 2009). Our best-fit photosphere models show some slight differences with the B9V - A1V spectra type derived by Currie *et al.* (2017) from optical spectra. In either case, the Tycho survey V-K = 0.205 photometric color for HD36546 is consistent with an unreddened ~A1V, both optical and near-IR spectra are inconsistent with spectral types earlier than A0, and neither solution affects our detection of excess NIR flux from circumstellar dust surrounding the star.

Adding in AKARI, WISE, and Herschel 3 - 160 μm photometry, we find a combined SED dominated by stellar photospheric emission at short wavelengths and an excess in the 20 - 160 μm range due to cold outer system dust. The long wavelength excess is normal, as we see it in more than half of all NIRDS + Spitzer IRS combined SEDs (Lisse *et al.* 2018). The total flux excess for the cold (T ~135K) outer dust is $L_{cold\ excess}/L_* \sim 4 \times 10^{-3}$, consistent with previous reports (McDonald *et al.* 2012; Wu *et al.* 2013; Liu *et al.* 2014; Cotten & Song 2016).

The SpeX spectrum + AKARI/WISE/Herschel photometry suggested an unusual spectral behavior: a secondary local SED maximum at ~8 μm, not the usual 9 - 11 μm created by silicaceous dust species. Our new BASS 3-13 μm spectrum confirms this. Figure 2 shows that while the excess observed in the 3 - 12 μm AKARI/WISE photometry, with maximum near 8 μm, can be fit with a blackbody model at T ~ 570K (dashed cyan curve), the BASS spectra and AKARI 8.8 μm photometry cannot. The BASS excess is similar to the unusual T-ReCS spectra reported for HD131499 and HD121191 by Melis *et al.* (2013), with their apparent peaks shortward of 8 μm (Fig. 2). Searching through the Spitzer/IRS 600+ disk, 5 - 35 μm survey database (Mittal *et al.* 2015) for more proof of this unusual excess behavior, we found a very good match with the excess spectrum of HD148657. This Spitzer flux excess has been scaled and overlaid in Figure 2; the match to HD36546's SED is clear, as is the non-blackbody shape of HD148657's 5 - 12 μm excess.

## 4. Discussion

Comparing the HD36546 SED to that of other well-studied A-star circumstellar disk systems showing excesses without narrowband gas line features (Chen *et al.* 2014, Mittal *et al.* 2015), we find some similarities and differences. In the outer belt dominated systems, seen in roughly half our NIRDS observations, there is a large peak at long wavelengths due to cold 50 - 150K dust (Lisse *et al.*

2018). In systems with strong warm rocky dust emission, seen in ~15% of our NIRDS systems (Dore *et al.* 2017; Lisse *et al.* 2015, 2018), strongly rising near-infrared excesses associated with 250 - 500K warm dust and prominent mid-infrared maxima at 9 - 11 μm are found. Both of these contributions can exist, e.g. in the A6V β Pic SED system (Chen *et al.* 2007). But for HD36546, while we find the long wavelength excess due to thermal emission from outer system dust, the near-infrared excess is only slowly increasing, if at all, towards longer wavelengths, and a secondary minor mid-infrared peak is found in the 7-8 μm range, clearly different from the normal 9-11 μm silicate dust maxima or the rarer ~9 μm silica dust maxima (e.g. A7V HD172555; Lisse et al. 2009).

We currently know of 3 other systems with matching spectral behavior; all are A0-A8V and young. The NIR excess for HD131488 (A3V, ~10 Myr) in our SpeX measurements is even larger in magnitude and extends down to ~3.2 μm (Fig. 1), suggesting a larger population of unusual inner system dust in this system. The disk morphologies for HD131488, as revealed by new GPI data, is similar as well. HD121191 (A8V, ~10 Myr) also demonstrates a warm and cold dust excess, although at lower SNR in the MIR (Melis *et al.* 2013) while imagery by Herschel (Vican *et al.* 2016) finds an extended cold dust disk for the system. A search through the 600+ Spitzer/IRS disk spectral survey (Chen *et al.* 2014, Jang-Condell *et al.* 2015, Mittal *et al.* 2015) turned up an excellent match to the HD36546 SED excess for the young A-star system, HD148657, with a similar broad maxima over the photosphere in the 5 - 13 μm region and a large cold (~135K) dust maximum (Fig. 2). The warm flux excess for HD148657 shows a maximum in the 7-8 um region dominated by C-C, C-Si, and C-H vibrational stretching modes, and minima in the ~6 μm water H-O-H bending mode region and the 9 - 12 μm Si-O stretching region.

What could be the cause of an apparent relative maximum at 7-8 μm in the HD36546, HD131488, HD121191, and HD148657 SEDs? A-stars illuminate and irradiate their massive primordial disks very strongly, so they are expected to evince rapid disk evolution (Melis *et al.* 2013, Chen *et al.* 2014, Mittal *et al.* 2015) and a high frequency of detected debris disks (Wu *et al.* 2013). The 7-8 μm emission peak can be sourced by species like carbonates, SiC, or aliphatic/aromatic hydrocarbons, all carbon-rich. By reference to our own solar system, if the HD36546 circumstellar material were located here, the dust population producing the warm, ~570K excess would be located inside the orbit of Mercury, while the cold ~135 K dust would be located in the outer regions of our asteroid belt.

Putting these ideas together, we can synthesize a few likely possibilities for the observed SED's: **(1)** The relative maximum at 7-8 μm could be coincidental, and each system could simply be hosting an inner belt at a few AU from the primary star that is currently producing a large amount of carbon-rich dust via collisional grinding and/or aggregation of carbon-planetisimals. **(2)** It could be that a giant impact has recently occurred involving a carbonaceous parent body (like a KBO or comet from the outer cold disk population) in the inner portions of these systems, producing carbon-rich dust warm enough to express a 6-9 μm emission feature. Or **(3)** it could be that we are finding emission pinned to the ~550-600 K sublimation range of common primitive dust species, like ferromagnesian sulfides, as the host A-star evaporates the last remnants of its primordial disk material, and the sublimation is also releasing hot carbonaceous material previously held in matrix.

A final possibility we mention for completeness' sake is **(4)** that the unusual A-star SEDs are composed of the typical photospheric + cold outer dust flux contributions plus "contamination" from background interstellar cirrus (c.f. Zubko, Dwek, & Arendt 2004, Flagley *et al.* 2006, Compiègne *et al.* 2011). After all, HD36546 was not listed in the original IRAS catalogue of Vega-like stars (Aumann 1985) precisely because it was confused with the highly structured Taurus dark cloud complex. However, this would seem to require an unlikely large cirrus contribution for HD36546's 0.3 Jy @ 8 μm flux that is not removed by the usual chopping/nodding performed in MIR observations, is at odds with the small value of E(B-V) = 0.01 found in our photospheric fits, and is not consistent with the compact MIR point sources seen in AKARI and WISE imagery. It would also require that the other 3 systems we have identified as similar, HD131488, HD121191, and HD148657, all be "contaminated" as well.

Currie *et al.* (2017), using forward modeling, found an imaged HD36546 dust disk with radius 85 ± 10 AU (FWHM). Assuming simple blackbody behavior and a total luminosity $L_* = 20\ L_\odot$ for HD36546, this dust should radiate as blackbodies with temperatures between 69K (75 AU) and 61K (95 AU). Blackbody dust at ~135K, as found in the longest wavelength source peaking at ~40 μm of our HD36546 SED (Fig. 2) would have to be located at ~20 AU from a 20 $L_\odot$ primary. Allowing for the superheat ratio $T_{obs}/T_{BB}$ = 100K/71K = 1.41 found by NIRDS for dust in the HR 4796A (A0.5V) ring at ~75 AU from its primary (Lisse *et al.* 2017) moves this location out to ~40 AU, closer to the Curie *et al.* estimate. The remaining discrepancy between the spectral and imaging radial distance estimates might be due to our poor current characterization of the MIR-FIR SED peak; to our lack of understanding of the superheat behavior of the HD 36546 outer dust (values of $T_{obs}/T_{BB}$ out to 5-15 have been reported by Rodriguez & Zuckerman 2012 and Vican *et al.* 2016) ; or to the different sensitivities of scattered light ($\propto$ surface area*albedo) vs. thermal emission ($\propto$ surface area*emissivity*$T^4$) measurements of circumstellar dust in a thick disk, which will weight closer-in dust thermal emission most highly (c.f. Watson *et al.* 2009 and references therein).

Using the T~570K fit to the continuum underlying the warm SED excesses in Fig. 2, we find the dust component creating the ~8 μm peaked emission would have to lie somewhere near 1.1 AU for pure blackbody behavior, or near 2.2 AU for superheated grains with $T_{obs}/T_{BB}$ = 1.41. Either estimate is well within the Currie *et al.* coronographic masking region, so no direct comparison is possible.

## 5.     Conclusions

**(1) The HD 36546 primary is a nearly unextincted early A star, not a late B star.** Our modeling favors A1.5V +/-1 subclass, which is slightly later than the A0V classification found from optical spectra (Currie et al. 2017). Both optical and near-IR spectra rule out spectral types earlier than A0. The best fit photospheric models require very low or no extinction, implying that HD36546 is in front of the 1–2 Myr old Taurus-Auriga star-forming region as seen from Earth, and that there is little to no in-system extinction of the primary star.

**(2) There are no detected** 2.1 - 2.5 μm CO **emission lines** due to fluorescing circumstellar gas nor HBrγ lines due to accretion, nor HeI 1.083 μm, FeII 1.256/1.644 μm, and $H_2$ S(1) 1-0 2.121 μm emission lines due to strong wind outflow. No emission line due to neutral PAHs are present at 3.29 μm.

**(3)** From 0.7 - 4.0 μm, **the SED is not "flat" or "filled-in" by inner disk hot dust** as in YSOs or transition disks. There are the beginnings of **warm dust excess flux in the 4.5 - 5.0 μm region**, agreeing with archival JHKLM photometry of the system (triangles in Fig. 2) showing a rise in M-band and WISE W2.

**(4) There are indications of an unusual local maximum in the SED at 6-9 μm.** A similar feature was seen in the A3V, ~12 Myr old debris disk system HD131488 and at lower SNR for the similarly bright HD121191 system (A8V, ~10 Myr) by Melis *et al.* (2013) using T-ReCS. Mittal *et al.'s* 2015 Spitzer/IRS disk survey contains another good spectral match the HD148657 (A7V,12 Myr) system. The only good laboratory analogue matches producing emission peaks in this wavelength range - organics, carbonates, SiC, PAHs - are all carbon-rich.

**(5)** If the observed SED at long wavelengths is due to thermal emission from 2 populations of dust with temperatures of ~570 K and ~135 K, respectively, then these two **populations of dust are densest and hottest at 1.1 - 2.2 AU and 20 - 40 AU from the ~20 $L_\odot$ primary, likely at the inner edges of their corresponding belts**. The ~1.6 AU dust may be the last remnant of the system's primordial disk, or due to sourcing from an aggregating or grinding asteroid belt in an old transition disk/very young debris disk. The ~30 AU outer dust population does not match up very well with the ~85 AU radial extent for the system's disk found by Currie *et al.* (2013) using H-band scattered light imaging, unless this outer belt is very wide and is dominated in the thermal IR by highly superheated dust at the belt's inner edge closest to the primary.

**Summary**. This system is a nearly mature transition disk, without the gas found in primordial and young transition disks. If it is of 3-10 Myr age, as argued by Currie *et al.* (2017), we favor the higher end of this range as the most likely due to the gas poor, nearly **cleared-out nature** of the disk. There are 2 belts of material making up the system's disk, a hot inner one at a few AU from the star, and the cold outer one imaged by Currie *et al.* (2017).

## 6.    Acknowledgements

the National Aeronautics and Space Administration, Science Mission Directorate, Planetary Astronomy Program. Our NIRDS observations take advantage of and add to the SpeX spectral library of ~200 cool FGKM stars (Rayner *et al.* 2009), and we are deeply indebted to J. Rayner for building SpeX and for providing his observing time and expertise to this project, and to C Chen and R. Patel for many useful discussions. C. Lisse would also like to gratefully acknowledge the support and input of the NASA Nexus for Exoplanet System Science (NExSS) research coordination network sponsored by NASA's Science Mission Directorate.## 7.  References

Aumann, H.H. 1985, *PASP* **97**, 88

Chen, C. H., *et al.* 2014. *Astrophys J. Suppl* **211**, 25

Compiègne, M. *et al.* 2011, *Astron. Astrophys.* **525**, A103

Connelley, M. S. and T. P. Greene 2010. *Astron. J.* **140**, 1214

Connelley, M. S. and T. P. Greene 2014. *Astron. J.* **147**, 125

Cotten, T.H. & Song, I. 2016, *Astrophys J. Suppl* **225**, 15

Currie, T. *et al.* 2017, "Subaru/SCExAO First-Light Direct Imaging of a Young Debris Disk around HD 36546", eprint arXiv:1701.02314

Cushing, M.C., Vacca, W.D., and Rayner, J.T. 2004, *PASP* **116**, 362

Dore, O., *et al.* 2016. "Science Impacts of the SPHEREx All-Sky Optical to Near-Infrared Spectral Survey", https://arxiv.org/pdf/1606.07039.pdf

Flagey, N. *et al.* 2006, *Astron. Astrophys.* **453**, 969

Hackwell, J.A. *et al.* 1990, *SPIE* **1235**, 171

Jang-Condell, H. *et al.* 2015, *Astrophys J.* **808**, 167

Kenyon, S. J., Gomez, M., Whitney, B. A., 2008, Handbook of Star Forming Regions, Volume I: The Northern Sky ASP Monograph Publications, Vol. 4. Edited by Bo Reipurth, p.405

Lisse, C.M. *et al.* 2012, *Astrophys J.* **747**, 93

Lisse, C.M. *et al.* 2013, *AAS Meeting* **#221**, id.325.04 2015

Lisse, C.M., Sitko, M.L., Marengo, M. 2015, *Astrophys J.* Lett. **815**, L27

Lisse, C.M., Sitko, M.L., Marengo, M., Vervack, R.J.,,Fernandez, Y.R., & Mittal, T. 2017, *"IRTF/SPEX Observations of the HR 4796A Cometary Ring System"*, *Astron. J.* (in press)

Lisse, C.M.*et al.* 2018. *"First Results From the IRTF/SPEX Near-Infrared Disk Survey (NIRDS)"*, *Astrophys J.* (in preparation)

Liu, Q., Wang, T., & Jiang, P. 2014, *Astron. J.* **148**, 3

Luhman, K. L., Mamajek, E. E., Allen, P. R., Cruz, K. 2009, *Astrophys J.* **703**, 399
11

McDonald, I., Zijlstra, A. A., & Boyer, M. L. 2012, *MNRAS* **427**, 343

Melis, C. *et al.* 2013, *Astrophys J.* **778**, 12

Mittal, T. *et al.* 2015, *Astrophys J.* **798**, 87

Padgett, D. 2011. "New Warm Debris Disks from WISE and Herschel" AAS/ESS meeting #2, id.38.06

Pecaut, M. & Mamajek, E. E., 2013, *Astrophys J. Suppl* **208**, 9

Pickles, A. & Depagne, É. 2010, *PASP* **122**, 1437

Rayner, J.T. *et al.* 2003, *PASP* **115**, 362

Rayner, J. T., M. C. Cushing, and W. D. Vacca 2009, *Astrophys. J. Suppl* **185**, 289

Rodriguez, D. R. & Zuckerman, B. 2012, *Astrophys J.* **745**, 147

Torres, R.M.*et al.* 2009, *Astrophys J.* 698, 242

Vacca, W.D., Cushing, M.C., and Rayner, J.T. 2003, *PASP* **115**, 389

Vacca, W.D., Cushing, M.C., and Rayner, J.T. 2004, *PASP* **116**, 352

van Leeuwen, F., 2007, *Astron. Astrophys.* **474**, 653

Vican, L. *et al.* 2016, *Astrophys J.* **833**, 263

Watson, D.M. *et al.* 2009, *Astrophys J. Suppl* **180**, 84

Wu, C.-J., Wu, H., Lam, M.-I., *et al.*, 2013, *Astrophys J. Suppl* **208**, 29

Zubko, V., Dwek, E., & Arendt, R.G. 2004, *Astrophys J. Suppl.* **152**, 211

# 8. Figures

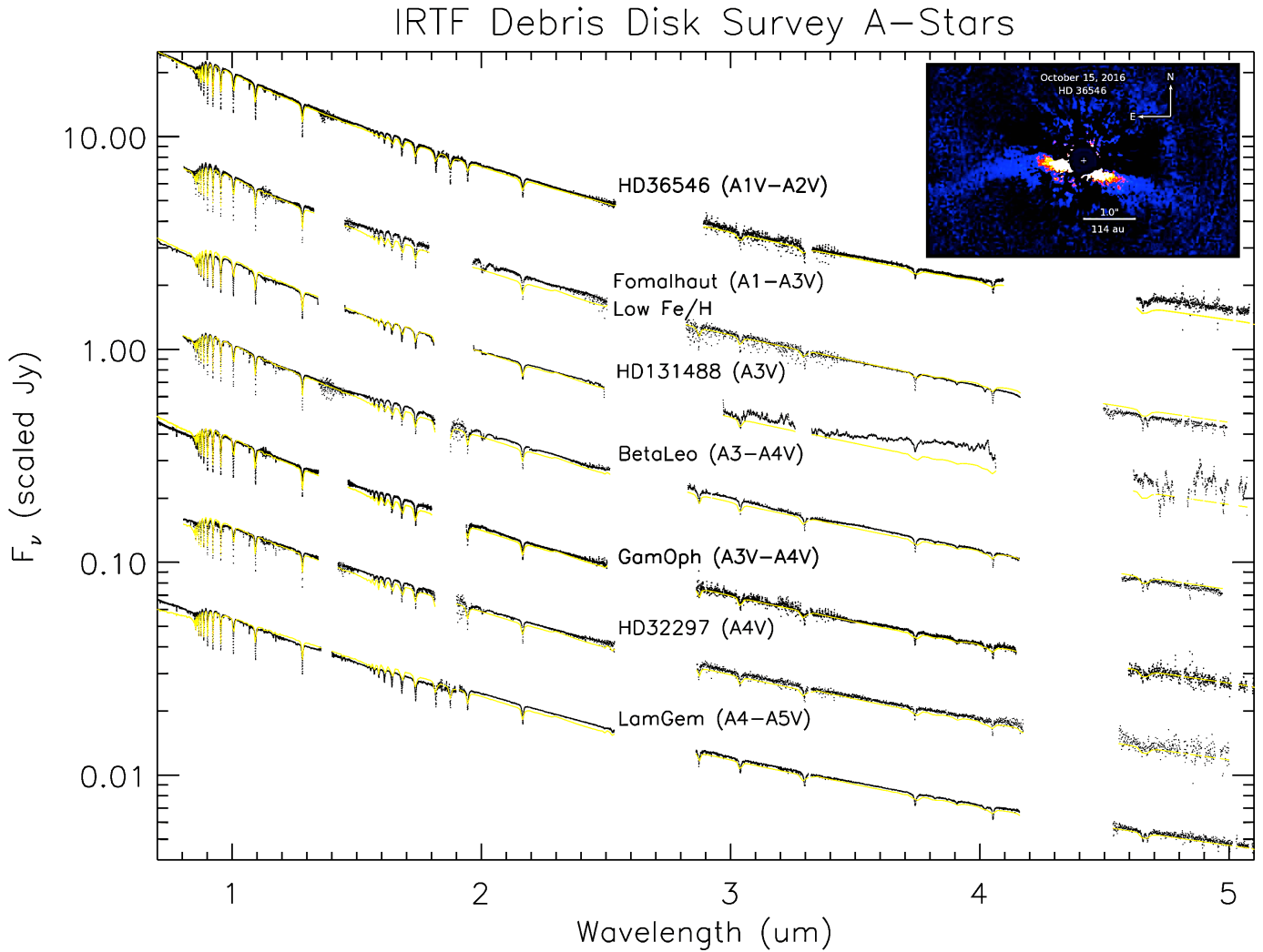

**Figure 1: Main sequence early A-stars with reported circumstellar material observed from the NASA IRTF under the NIRDS program.** R~1000, 0.7-5.0 μm SpeX data are in black and photospheric spectral models are in gold. The typical spectral behavior for our program systems matches the photospheric model well from 0.7 - 1.3 μm, exceeds it slightly (if at all) from 1.3 - 3 μm, then matches it again from 3 - 5 μm. HD 131488 shows the typical growing exponential behavior seen for other NIRDS stars (e.g., HD113766, HD15407A, HD23514; Lisse *et al.* 2015) heavily dominated past 3 μm by warm (~300K) thermal dust emission observed on the Wien law side of its SED. The spectral behavior of HD 36546 in the near-infrared matches A0.5-A2.5V photospheric emission models very well from 0.7 to 4.1, but from 4.5 to 5.0 μm an excess flux above photospheric due to circumstellar material is found. *Inset*: Subaru SCExAO/HiCIAO H-band image of the HD36546 disk from Currie *et al.* (2017).

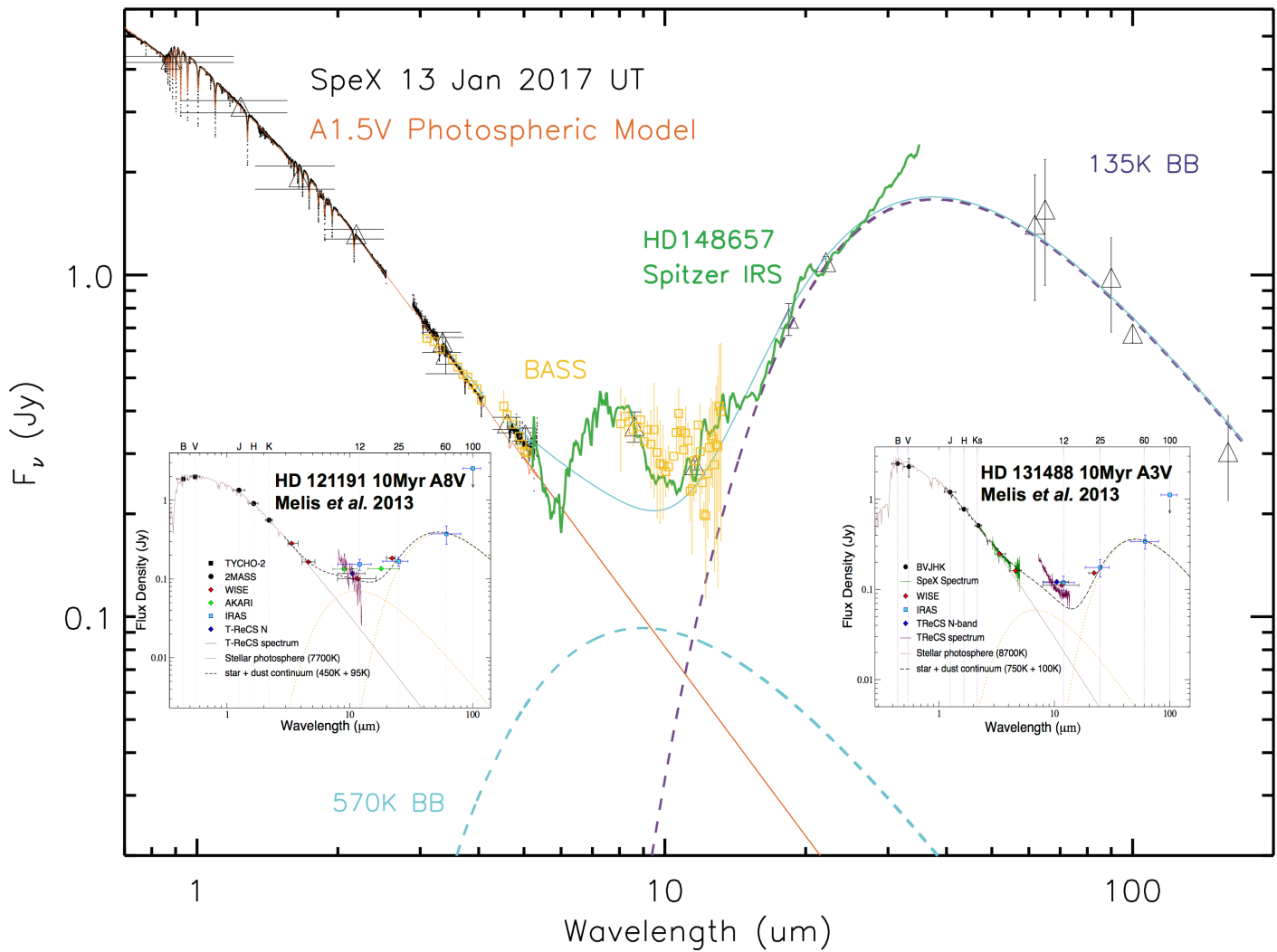

**Figure 2: 0.7 - 160 μm SED of HD36546** created by adding together the 13 Jan 2017 UT SpeX data (black points) with Vizier archival Tycho/2MASS/WISE/AKARI/Herschel photometry of the system (triangles). Only photometry with published error values has been included in the SED; the error bars presented in the figure are ± 2σ. Overlain in orange is our best-fit NextGen E(B-V) = 0.01, $T_{eff}$ = 8920 ± 80 K, [Fe/H] = 0.0, log g = 4.50 ± 0.25 photospheric model, corresponding to an A1.5V star. The positive deviation from photospheric emission seen in the SpeX data at 4.5 - 5.0 μm is echoed in the M-band photometry. A large excess flux at 18 to > 30 μm seen in the other NIRDS A-star systems, due to cold outer system dust in the system's Centaur - to Kuiper Belt regions, is present. Also present is an unusual local maximum at 8.5 μm and minimum at 11 μm, possibly similar to the ones found in SpeX+T-ReCS spectra + archival 2MASS/WISE/IRAS photometry of the ~10 Myr old HD131488 and HD 121191 A-star systems (Melis *et al.* 2013, **insets**) and the Spitzer/IRS excess spectrum of HD148657 (green curve). The cyan curve is a failed attempt to fit the 8-11 μm archival photometry + our SpeX spectrum simultaneously by adding in a extra warm dust blackbody component (see text). We conclude that the anomalous mid-IR flux is due to a non-black body source, such as a strong solid-state line or feature.